\documentclass[%
 reprint,
 superscriptaddress,
 amsmath,amssymb,
 aps,
 pre,
 showkeys,
 floatfix,
]{revtex4-2}

\usepackage{graphicx}       
\usepackage[caption=false]{subfig}
\usepackage{dcolumn}        
\usepackage{bm}             
\usepackage{hyperref}       
\usepackage{xcolor}         
\usepackage{physics}        
\usepackage{multirow}
\usepackage{makecell}       
\usepackage{array}          
\usepackage{mathtools}      

\iftrue
    \newcommand{\todo}[1]{{\color{red}[#1]}} 
    \newcommand{\question}[1]{{\color{blue}[#1]}}
\else
    \newcommand{\todo}[1]{}               
    \newcommand{\question}[1]{}
\fi

\newcommand{\deri}[2]{\frac{\partial #1}{\partial #2}}

\renewcommand{\vec}[1]{\boldsymbol{#1}}
\newcommand{\mat}[1]{\mathbf{#1}}
\DeclareMathSymbol{\shortminus}{\mathbin}{AMSa}{"39}

\newcommand{\boltz}{k_{\text{B}}}
\newcommand{\bohr}{a_{\text{B}}}

\begin{document}

\preprint{APS/123-QED}

\title{Modelling of warm dense hydrogen via explicit real time electron dynamics:\\ Electron transport properties}

\author{Pontus~Svensson}
\email{pontus.svensson@physics.ox.ac.uk}
\affiliation{Department of Physics, University of Oxford, Parks Road, Oxford OX1 3PU, UK}%

\author{Patrick~Hollebon}
\affiliation{AWE, Aldermaston, Reading, Berkshire RG7 4PR, UK}%

\author{Daniel~Plummer}
\affiliation{Department of Physics, University of Oxford, Parks Road, Oxford OX1 3PU, UK}%

\author{Sam~M.~Vinko}
\affiliation{Department of Physics, University of Oxford, Parks Road, Oxford OX1 3PU, UK}%
\affiliation{Central Laser Facility, STFC Rutherford Appleton Laboratory, Didcot OX11 0QX, UK}%

\author{Gianluca~Gregori}
\affiliation{Department of Physics, University of Oxford, Parks Road, Oxford OX1 3PU, UK}%

\date{\today}

\begin{abstract}
    We extract electron transport properties from atomistic simulations of a two-component plasma, by mapping the long-wavelength behaviour to a two-fluid model. The mapping procedure is performed via Markov Chain Monte Carlo sampling over multiple spectra simultaneously. The free-electron dynamic structure factor and its properties have been investigated in the hydrodynamic formulation to justify its application to the long-wavelength behaviour of warm dense matter. We have applied this method to warm dense hydrogen modelled with wave packet molecular dynamics, and showed that the inferred electron transport properties are in agreement with a variety of reference calculations, except for the electron viscosity, where a substantive decrease is observed when compared to classical models.
\end{abstract}

\maketitle

\section{Introduction}\label{sec:intro}
Dense and highly non-ideal plasmas can be readily found in astrophysical objects~\cite{saumon1992role,chabrier1993quantum,guillot1999interiors,fortney2010interior} and during inertial confinement fusion~\cite{nuckolls1972laser,abu2022lawson} experiments~\cite{hu2018review}. Warm dense matter (WDM) is one such regime, characterised by strong ion correlations and partial electron degeneracy~\cite{bonitz2020ab}, a challenging combination of attributes to model theoretically and uncertainties remain for many basic properties~\cite{gaffney2018review}. Transport coefficients are especially uncertain, and two recent workshops have demonstrated orders of magnitude discrepancies between models~\cite{grabowski2020review,stanek2024review}.

These dense plasma states can now be created in the laboratory using high-power lasers~\cite{kritcher2008ultrafast,falk2014equation,levy2015creation,fletcher2015ultrabright,falk2018experimental,smith2018equation}, however, due to an interplay of complicating factors -- e.g. the transient realisation of the state and a mixture of thermodynamic conditions -- uncertainties remain in the transport properties, which influence the accuracy of predictive simulations~\cite{weber2014inhibition,hu2014first,hu2015impact,vold2017plasma}.

First principle models based on \textit{Molecular Dynamics} (MD) or \textit{Monte Carlo} (MC) approaches fully account for interactions and correlations, but are limited in the physical size modelled. Methods to extract the hydrodynamic properties from first principle computations are therefore needed. For example, equation of state data has been tabulated and parameterised, e.g.\ by the widely used SESAME~\cite{lyon1978sesame} and FPEOS~\cite{militzer2021first} tables. Here we discuss a method for extracting electron transport properties from MD simulations of two-component plasmas. The method consists of mapping an analytical hydrodynamic model to the long-wavelength behaviour of the simulated dynamic structure factor via Bayesian parameter estimation~\cite{von2011bayesian}. The method allows for validation of the hydrodynamic model in question and the estimation of errors and correlations between inferred properties. For the atomistic model, we consider dynamic structure factors computed from wave packet molecular dynamics (WPMD)~\cite{klakow1994hydrogen,feldmeier2000molecular,grabowski2014review}, which we previously discussed in Ref.~\cite{svensson2024modelling}, and is henceforth referred to as paper I. The presented method is general and can be applied to any modelling technique or even experimental data with access to the electron dynamic structure factor at sufficiently long wavelengths.

The remainder of the manuscript is structured as follows. Section \ref{sec:hydro} describes a two-fluid model that predicts the main long-wavelength characteristics of an electron fluid interacting with ions. The requirements for appropriately applying the hydrodynamic model to MD data are discussed in section \ref{sec:system}. The Bayesian inference model is discussed and applied to MD data in sections \ref{sec:MCMC} and \ref{sec:inference} respectively. Section \ref{sec:conclusion} summarises the conclusions.

\section{Hydrodynamic model}\label{sec:hydro}
The hydrodynamic model for the electron-electron dynamic structure factor, $S_{ee}(\vec{k}, \omega)$, must capture plasma oscillations which are primarily damped by the interaction with ions. Therefore, a two-fluid model is examined which has collisional momentum and energy transfer between electrons and ions, in combination with Poisson's equation for the large scale electric fields. Schmidt~\textit{et al.}~\cite{schmidt2012quantum} considered a quantum hydrodynamic formulation for electrons in a single fluid treatment, however, the correct limiting behaviour of the plasmon feature was not obtained due to a lack of electron-ion collisions. We therefore consider a two-fluid formulation, but limit ourselves to a classical hydrodynamic formulation, which is further motivated in section \ref{sec:system}. Specifically, the fluid equations of interest are for electrons ($\mu = e$, $\Bar{\mu} = i$) and ions ($\mu = i$, $\Bar{\mu} = e$) respectively, 
\begin{subequations}
\begin{equation}
    \frac{D \rho_\mu}{D t} + \rho_\mu \vec{\nabla} \cdot \vec{v}_\mu = 0,
    \label{eq:fluid:mass}
\end{equation}
\begin{equation}
    \rho_\mu \frac{D \vec{v}_\mu}{D t} + \vec{\nabla} \cdot \mat{P}_{\mu} = - \frac{q_{\mu}}{m_{\mu}}\rho_{\mu} \vec{\nabla} \phi - \vec{R}_{\mu\Bar{\mu}}\left(\rho_{\mu}, \vec{v}_\mu - \vec{v}_{\Bar{\mu}} \right),
    \label{eq:fluid:momentum}
\end{equation}
\begin{equation}
    \begin{aligned}
        \rho_\mu \frac{D \epsilon_{\mu}}{D t} + \mat{P}_\mu \colon \vec{\nabla}\vec{v}_\mu + \vec{\nabla} \cdot \vec{q}_{\mu} = &-\frac{q_\mu}{m_\mu} \rho_\mu \vec{v}_\mu \cdot \vec{\nabla}\phi\\
        &- g \left( T_{\mu} - T_{\Bar{\mu}} \right),
        \label{eq:fluid:energy}
    \end{aligned}
\end{equation}%
\label{eq:fluid}%
\end{subequations}%
which in convective form describes the conservation of mass, momentum and energy. In the above, $m_\mu$, $q_\mu$, $\rho_\mu$, $\vec{v}_{\mu}$ and $\epsilon_{\mu}$ are the particle mass, particle charge, mass density, (fluid) velocity and internal energy (per unit mass), and 
\begin{equation}
    \frac{D}{D t} \equiv \deri{}{t} + \vec{v}_\mu \cdot \vec{\nabla},
\end{equation}
is the material derivative.

The constitutive relations for the pressure tensor are chosen to be of Navier–Stokes type~\cite{boon1991molecular,currie1993fundamental,hansen1993theory},
\begin{equation}
    \begin{aligned}
        \mat{P}_{\mu} &= \left[p_\mu - \zeta_\mu \left(\vec{\nabla} \cdot \vec{v}_\mu \right)\right] \mat{I}\\
        &- \eta_{\mu} \Big[ \vec{\nabla}\vec{v}_\mu + \left( \vec{\nabla}\vec{v}_\mu \right)^{\intercal} - \frac{2}{3} \left(\vec{\nabla} \cdot \vec{v}_\mu \right) \mat{I} \Big],
    \end{aligned}
    \label{eq:pressure_tensor}
\end{equation}
where $p_{\mu}$ is the (thermodynamic) pressure and $\eta_{\mu}$ and $\zeta_{\mu}$ as shear and bulk viscosity respectively. The heat flux is described by Fourier's law~\cite{boon1991molecular,hansen1993theory,currie1993fundamental}, 
\begin{equation}
    \vec{q}_\mu = - \kappa_\mu \vec{\nabla} T_\mu,
    \label{eq:heat}
\end{equation}
where $\kappa_\mu$ is the thermal conductivity.

Momentum transfer via collisions is modelled by
\begin{equation}
    \vec{R}_{\mu\Bar{\mu}}\left(\rho_{\mu}, \Delta \vec{v} \right) = \int_{-\infty}^{\infty}d\tau\; \rho_{\mu}(\tau) r_{\mu\Bar{\mu}}(\tau)\, \Delta \vec{v}(t - \tau),
    \label{eq:collisions}
\end{equation}
where momentum conservation is satisfied if $\rho_{e} r_{ei} = \rho_{i} r_{ie}$. The Fourier transform of $r_{\mu\Bar{\mu}}$ is the dynamic collision frequency $\nu_{\mu\Bar{\mu}}(\omega)$ which must satisfy Kramers–Kronig relations due to causality constraints. The motivation for this exact form will be apparent in section~\ref{sec:hydro:lim}. Energy exchange, in equation \eqref{eq:fluid:energy}, is described in a two-temperature model with temperature relaxation rate $g$~\cite{glosli2008molecular,stanek2024review}, commonly used in combination with Fourier heat flow, e.g. Refs.~\cite{anisimov1974electron,ivanov2003combined,jiang2008plasma}, and is a special case of a more general heat transport model~\cite{qiu1993heat}.

Lastly, the system of equations \eqref{eq:fluid} is closed by Poisson's equation for the electrostatic potential $\phi$, 
\begin{equation}
    \varepsilon_0 \vec{\nabla}^2 \phi = - \left(\frac{q_i}{m_i} \rho_i + \frac{q_e}{m_e} \rho_e\right).
    \label{eq:poisson}
\end{equation}
where $\varepsilon_0$ is the vacuum permittivity. The inclusion of this form of electrostatic interactions in hydrodynamics is motivated for collisional systems~\cite{[][{ Section 4.4}]baus1980statistical}.

Using thermodynamic relations, the energy conservation equation \eqref{eq:fluid:energy} may be reformulated in terms of a local temperature $T_\mu$~\cite{boon1991molecular,bott2019thomson},
\begin{equation}
    \begin{aligned}
        \rho_\mu C_{V,\mu} \frac{D T_{\mu}}{D t} &=- \vec{\nabla} \cdot \vec{q}_{\mu} + \left( \zeta_{\mu} - \frac{2\eta_{\mu}}{3}\right) \left(\vec{\nabla}\cdot \vec{v}_\mu \right)^2\\
        &\hphantom{=}+ \eta_{\mu} \left[\vec{\nabla}\vec{v}_{\mu} + \left( \vec{\nabla}\vec{v}_{\mu}\right)^{\intercal} \right] \colon \vec{\nabla}\vec{v}_\mu\\
        &\hphantom{=}-\rho_\mu C_{V,\mu} \frac{\gamma_\mu - 1}{\alpha_{T,\mu}} \vec{\nabla}\cdot\vec{v}_\mu\\
        &\hphantom{=}-\frac{q_\mu}{m_\mu} \rho_\mu \vec{v}_\mu \cdot \vec{\nabla}\phi - g \left( T_{\mu} - T_{\Bar{\mu}} \right),
        \label{eq:temperature}
    \end{aligned}
\end{equation}
where $C_{V,\mu}$ is the specific heat a constant volume, $\gamma_\mu$ is the adiabatic index and $\alpha_{T,\mu}$ is the coefficient of thermal expansion. Furthermore, the pressure may be eliminated from the system through the application of~\cite{boon1991molecular,bott2019thomson},
\begin{equation}
    \vec{\nabla}p_\mu = \frac{c_{s,\mu}^2}{\gamma_\mu} \left( \vec{\nabla}\rho_\mu + \rho_\mu \alpha_{T,\mu} \vec{\nabla}T_\mu \right),
    \label{eq:pressure_relation}
\end{equation}
which introduces ${c_{s,\mu}^2 = (\partial p_\mu / \partial \rho_\mu)_{s_\mu}}$ which for classical fluids is the adiabatic sound speed.

Equations \eqref{eq:fluid:mass}, \eqref{eq:fluid:momentum} and \eqref{eq:temperature} will form the basis of our hydrodynamic analysis, using \eqref{eq:pressure_tensor}, \eqref{eq:heat}, \eqref{eq:collisions}, \eqref{eq:poisson} and \eqref{eq:pressure_relation} to close the system. The transport coefficients are left unspecified and shall be extracted by comparison to atomistic simulations in section \ref{sec:inference}. This gives a rather general formulation which is, however, limited to small wave numbers $\vec{k}$ and low frequencies $\omega$. Theories that generalise to finite $\vec{k}$ and $\omega$, so-called viscoelastic theories~\cite{lovesey1984theory,boon1991molecular,hansen1993theory}, introduce additional \textit{a priori} unknown coefficients to be inferred. Thus, to avoid further complication, we restrict ourselves to only the hydrodynamic regime.

\subsection{Linearised hydrodynamic equations}\label{sec:hydro:lin}
To investigate correlation spectrum of small amplitude density-density fluctuations, the fluid system is linearised around a static and homogeneous equilibrium. Due to the large mass ratio $m_i / m_e$, motions develop on two distinct time scales, the comparatively \textit{slow} ion and screening dynamics, and the \textit{fast} plasma oscillations. Specifically, we consider the perturbations,
\begin{equation}
    \begin{aligned}
    \rho_{e} &= \rho_{0, e}  + \delta \Bar{\rho}_e + \delta \rho_{e},\\
    \vec{v}_{e} &= \delta \Bar{\vec{v}}_{e} + \delta \vec{v}_{e},\\ 
    T_{e} &= T_{0, e} + \delta \Bar{T}_{e} + \delta T_{e},
    \end{aligned}
    \quad
    \begin{aligned}
        \rho_{i} &= \rho_{0, i} + \delta \Bar{\rho}_{i,0},\\
        \vec{v}_{i} &= \delta \Bar{\vec{v}}_{i},\\ 
        T_{i} &= T_{0, i} + \delta \Bar{T}_{i},
    \end{aligned}
\end{equation}
where $0$ index represents the equilibrium properties, the ''bar'' represents perturbations on slow time scales and the remaining perturbations are fast. Ions only exhibit slow perturbations. Charge neutrality requires $q_i \rho_{0,i} / m_i = -q_e \rho_{0,e} / m_e$ and temperature equilibrium is assumed $T_{0,i} = T_{0,e}$. The separation into slow and fast dynamics directly corresponds to the screening cloud and free electron terms in the Chihara decomposition~\cite{chihara1987difference,chihara2000interaction}, which is often used to interpret electron structure factors. For now, we are mainly concerned with free electron dynamics (fast perturbations), but the slow dynamics may also be solved to describe the electron screening of ions and ion-acoustic waves.

To isolate the fast dynamics, the electron equations are time averaged over time scales that are long with respect both the inverse (electron) plasma frequency $\omega_p^{-1}$ and the time scale of thermal electron motion. Over these longer time scales fast perturbations are assumed to average out and subtracting the result from the original equations isolates the fast perturbations to linear order,
\begin{subequations}
\begin{equation}
    \deri{\delta \rho_e}{t} + \rho_{0, e} \vec{\nabla} \cdot \delta \vec{v}_{e} = 0,
    \label{eq:lin:density}
\end{equation}
\begin{equation}
    \begin{aligned}
        \rho_{0, e} \deri{\delta \vec{v}_e}{t} = &- \frac{c_{s, e}^2}{\gamma_{e}} \left( \vec{\nabla} \delta \rho_{e} + \rho_{0, e} \alpha_{T, e} \vec{\nabla} \delta T_{e, 0} \right)\\
        &+ \eta_{e} \vec{\nabla}^2 \delta \vec{v}_e + (\zeta_{e} + \eta_e/3) \vec{\nabla}\vec{\nabla} \cdot \delta \vec{v}_{e}\\
        &-\frac{q_e}{m_e}\rho_{0, e} \vec{\nabla} \delta \phi - \vec{R}_{ei}\left(\rho_{0, e}, \delta \vec{v}_e \right),
    \end{aligned}
    \label{eq:lin:momentum}
\end{equation}
\begin{equation}
    \begin{aligned}
        \rho_{0, e} C_{V, e} \deri{\delta T_{e}}{t} = &- \rho_{0, e} C_{V, e} \frac{\gamma_{e} - 1}{\alpha_{T, e}} \vec{\nabla} \cdot \delta\vec{v}_e\\
        &+ \kappa_{e} \vec{\nabla}^2 \delta T_{e} - g \delta T_{e},
    \end{aligned}
    \label{eq:lin:temperature}
\end{equation}
\begin{equation}
    \vec{\nabla}^2 \delta \phi = -\frac{q_e}{m_e\varepsilon_0} \delta \rho_{e},
    \label{eq:lin:poissone}
\end{equation}%
\label{eq:lin}%
\end{subequations}%
and the transport coefficients are constant to linear order. In a higher order treatment -- an equivalent of \textit{Zakharov’s equations}~\cite{zakharov1972collapse} -- the slow and fast dynamics couple together, however, such treatment is beyond the scope of the current manuscript.

The density \eqref{eq:lin:density} and temperature \eqref{eq:lin:temperature} equations only couple to the divergence of the perturbed velocity field $\vec{\nabla}\cdot \delta \vec{v}_e$. Therefore, two velocity degrees of freedom -- the shear modes -- decouple from the density fluctuations~\cite{boon1991molecular} and only three modes will appear in the density dynamics, which will be shown to correspond to one entropy and two plasmon modes.

\subsection{Hydrodynamic fluctuations and dynamic structure factors}\label{sec:hydro:DSF}
The spectrum of density-density correlations is examined following the well-established methodology of Refs.~\cite{vieillefosse1975statistical,lovesey1984theory,boon1991molecular,hansen1993theory,schmidt2012quantum,bott2019thomson,mondal2024quantum}, where the dynamics of the perturbations are analysed through a spatial Fourier transform in combination with a temporal Laplace transform. Using the notation,
\begin{equation}
    \delta \Tilde{\vec{x}}^{\mu}_{\vec{k}}(s) = \int_{0}^{\infty}\! dt\, e^{-st} \int d^3\vec{r}\, e^{-i \vec{k}\cdot\vec{r}} \delta \vec{x}_{\mu}(\vec{r}, t),
\end{equation}
for a vector quantity $\delta \vec{x}_{\mu}$, the linearised system \eqref{eq:lin} may be written as,
\begin{equation}
    \left[s \mat{I} + \mat{D}_{e}(\vec{k}) \right]\, \delta \Tilde{\vec{y}}^{e}_{\vec{k}}(s) = \delta \vec{y}^{e}_{\vec{k}}(0),
    \label{eq:FL_lin}
\end{equation}
for combined perturbation variables $\delta \Tilde{\vec{y}}^{\mu}_{\vec{k}}(s)$ and initial conditions $\delta \vec{y}^{\mu}_{\vec{k}}(0)$, where 
\begin{widetext}
\begin{equation}
    \delta \Tilde{\vec{y}}^{\mu}_{\vec{k}}(s) = \Big[
        \delta \Tilde{\rho}^{\mu}_{\vec{k}}(s),\, i \rho_{0,\mu} \vec{k}\cdot\delta\Tilde{\vec{v}}^{\mu}_{\vec{k}}(s),\, \rho_{0, \mu} \delta \Tilde{T}^{\mu}_{\vec{k}}(s)\Big]^{\intercal},\quad \delta \vec{y}^{\mu}_{\vec{k}}(0) = \Big[
        \delta \rho^{\mu}_{\vec{k}}(0),\, i \rho_{0,\mu} \vec{k}\cdot\delta\vec{v}^{\mu}_{\vec{k}}(0) + i\vec{k}\cdot \vec{R}_{0,\vec{k}}/s,\, \rho_{0, \mu} \delta T^{\mu}_{\vec{k}}(0)\Big]^{\intercal}.
\end{equation}
The variables without a tilde are evaluated in the time domain and represent the initial condition. $\vec{R}_{0, \vec{k}}$ describes the contribution to the momentum transfer from $t < 0$. The linearised dynamics is completely determined by the hydrodynamic matrix,
\begin{equation}
    \mat{D}_{e}(\vec{k}) = \begin{bmatrix}
        0 & 1 & 0\\
        - \left( \omega_{p}^2 + \gamma_e^{-1} c_{s,e}^2 \vec{k}^2 \right) & \nu_{ei} + \nu_{L,e}\vec{k}^2 & -\alpha_{T,e} \gamma_{e}^{-1} c_{s,e}^2\vec{k}^2\\
        0 & \alpha_{T,e}^{-1}\left(\gamma_{e} - 1\right) & \gamma_{e}\left( \mathfrak{g}_e + \chi_{e} \vec{k}^2 \right)
    \end{bmatrix},
\end{equation}
and its eigenvalues describe the modes present in the system. The hydrodynamic matrix is defined in terms of the plasma frequency $\omega_p^2 = \rho_{0, e} q_e^2 / (m_e^2 \varepsilon_0)$, the longitudinal (kinematic) viscosity $\nu_{L,e} = \left(4\eta_e/3 + \zeta_e\right)/\rho_{0, e}$, thermal diffusivity $\chi_{e} = \kappa_e / (\rho_{0, e} \gamma_e C_{V, e})$ and the normalised heat transfer $\mathfrak{g}_{e} = g_e / (\rho_{0, e} \gamma_e C_{V, e})$. 
\end{widetext}

\begin{figure}
    \centering
    \includegraphics[width=\linewidth]{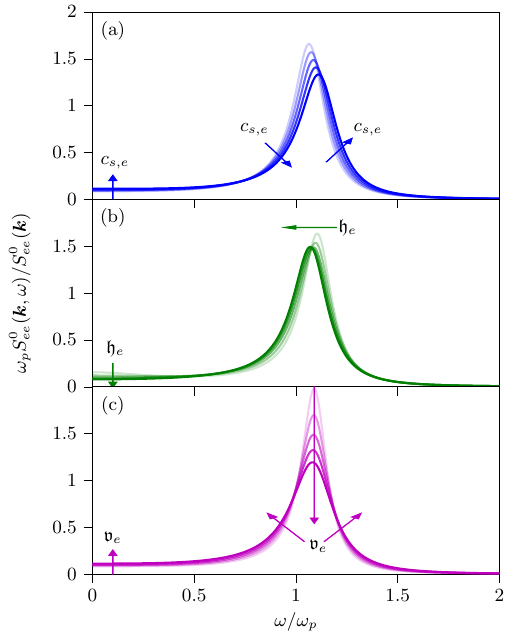}
    \caption{Schematic of the dependence on the coefficients (a) $c_{s,e}$, (b) $\mathfrak{h}_e$ and (c) (real part) $\mathfrak{v}_e$ on the hydrodynamic dynamic structure factor, equation \eqref{eq:hydro_DSF}. Arrows indicate the change as each variable is increased respectively.}
    \label{fig:understanding_transport_DSF}
\end{figure}

The (classical) electron dynamic structure factor, for a system of $N_e$ electrons, is
\begin{equation}
    S^{0}_{ee}(\vec{k}, \omega) = \frac{1}{2\pi} \int_{-\infty}^{\infty}\! dt\, \frac{e^{i \omega t}}{N_e m_e^2} \left\langle \delta \rho^{e}_{\vec{k}}(t) \delta \rho^{e*}_{\vec{k}}(0) \right\rangle,
\end{equation}
where $\langle \cdot \rangle$ is the thermal average. Using the transformed variables it may be evaluated as~\cite{boon1991molecular,hansen1993theory}, 
\begin{equation}
    \frac{S^{0}_{ee}(\vec{k}, \omega)}{S^{0}_{ee}(\vec{k})} = \frac{1}{\pi} \Re\left[ \lim_{\varepsilon \rightarrow 0} \frac{\langle \delta \Tilde{\rho}^{e}_{\vec{k}}(\varepsilon - i\omega) \delta \rho^{e*}_{\vec{k}}(0) \rangle}{\langle \delta \rho^{e}_{\vec{k}}(0) \delta \rho^{e*}_{\vec{k}}(0) \rangle} \right],
    \label{eq:fluctuations2spectrum}
\end{equation}
normalised by its integral $S^{0}_{ee}(\vec{k})$, the static structure factor. Under the common equilibrium assumption that the initial perturbations in density, velocity and temperature are statistically independent~\cite{lovesey1984theory,boon1991molecular,schmidt2012quantum,bott2019thomson,mondal2024quantum}, cross-correlations vanishes when equation \eqref{eq:FL_lin} is inverted and thermally averaged with the initial density perturbations,
\begin{equation}
    \langle \delta \Tilde{\rho}^{e}_{\vec{k}}(s) \delta \rho^{e*}_{\vec{k}}(0) \rangle = \left(s\mat{I} + \mat{D}(\vec{k}) \right)^{-1}_{11}\; \langle \delta \rho^{e}_{\vec{k}}(0) \delta \rho^{e*}_{\vec{k}}(0) \rangle.
\end{equation}
The final result is an explicit form for the dynamic structure factor,
\begin{equation}
    \frac{\pi S^{0}_{ee}(\vec{k}, \omega)}{S^{0}_{ee}(\vec{k})} = \Re\left[ \frac{s^2 + A(\vec{k}) s + B(\vec{k})}{s^3 + A(\vec{k}) s^2 + C(\vec{k}) s + D(\vec{k})} \right]_{\mathrlap{s = -i\omega}},
    \label{eq:hydro_DSF}
\end{equation}
where
\begin{subequations}
\begin{align}
    A(\vec{k}) &= \mathfrak{v}_{e}(\vec{k}, \omega) + \gamma_{e} \mathfrak{h}_{e}(\vec{k}),\\
    B(\vec{k}) &= \frac{\gamma_e - 1}{\gamma_e}c_{s,e}^2 \vec{k}^2 + \gamma_e \mathfrak{h}_{e}(\vec{k})\; \mathfrak{v}_{e}(\vec{k}, \omega),\\
    C(\vec{k}) &= \omega_p^2 + c_{s,e}^2\vec{k}^2 + \gamma_e \mathfrak{h}_{e}(\vec{k})\; \mathfrak{v}_{e}(\vec{k}, \omega),\\
    D(\vec{k}) &= \left(\gamma_e \omega_p^2 + c_{s,e}^2 \vec{k}^2\right) \mathfrak{h}_{e}(\vec{k}),%
\end{align}%
\label{eq:DSF_coefficents}%
\end{subequations}%
with $\mathfrak{v}_{\mu}(\vec{k}, \omega) = \left[\nu_{\mu\Bar{\mu}}(\omega) + \nu_{L,\mu} \vec{k}^2\right]$ and $\mathfrak{h}_{\mu}(\vec{k}) = \left[ \mathfrak{g}_\mu + \chi_\mu \vec{k}^2 \right]$. Note that, the structure factor is independent of the thermal expansion coefficient $\alpha_{T,e}$ which therefore can't be inferred from the scattering spectrum or the dynamic structure factor. The schematic dependence of $S_{ee}^{0}(\vec{k}, \omega)$ on the remaining coefficients $c_{s,e},\; \mathfrak{h}_{e}$ and $\mathfrak{v}_e$ is shown in figure \ref{fig:understanding_transport_DSF}. The position of the plasmon feature is mainly set by $c_{s,e}$ and $\mathfrak{v}_{e}$ sets its width. The entropy feature at $\omega = 0$ is primarily determined by $\mathfrak{h}_{e}$,  however, all coefficients modify multiple parts of the spectrum to some degree.

The form in equation \eqref{eq:hydro_DSF} is not written in terms of three Lorentzians, common for hydrodynamic forms, however, it still exhibits similar behaviour. The hydrodynamic form is obtained by expansion in small $\vec{k}$~\cite{schmidt2012theory}, however, such treatment does not directly extend to non-zero $\nu_{ei}$ and $\mathfrak{g}_e$, and will not be required here. 

\subsection{Limiting behaviours of the hydrodynamic formulation}\label{sec:hydro:lim}
In the limit where dissipative effects are small both via inter-species collisions as well as viscous and heat conductive effects, i.e. $\mathfrak{h}_{e} \ll \omega_p$ and $\mathfrak{v}_{e} \ll \omega_p$, the dispersion $|s\mat{I} + \mat{D}_e(\vec{k})| = 0$ can be evaluated analytically,
\begin{subequations}
\begin{align}
    s_0 &= -\left( \frac{\gamma_e \omega_p^2 + c_{s,e}^2\vec{k}^2}{\omega_Q^2} \right) \times \mathfrak{h}_{e},\label{eq:entropy_mode}\\
    s_{1,2} &= \pm i\omega_Q - (\gamma_e - 1) \frac{c_{s,e}^2 \vec{k}^2}{2\omega_Q^2} \mathfrak{h}_{e} - \frac{\mathfrak{v}_{e}(\vec{k}, \mp\omega_{Q})}{2},
\end{align}%
\label{eq:eigen_modes}%
\end{subequations}%
where $s = -\gamma - i\omega$ describes the resonance frequency $\omega$ and damping rate $\gamma$, and
\begin{equation}
    \omega_Q^2 = \omega_p^2 + c_{s,e}^2 \vec{k}^2.
    \label{eq:dipersion}
\end{equation}
Equation \eqref{eq:eigen_modes} show an entropy mode at $\omega = 0$ which is damped by heat transfer, and two plasma modes at $\omega \approx \pm \omega_Q$. The plasma modes are primarily damped by viscous drag and momentum transfer $\mathfrak{v}_{e}$, and not energy conduction or transfer $\mathfrak{h}_e$ because the plasma mode is fast $\omega_p \gg c_{s,e}\vec{k}$, as previously illustrated in figure \ref{fig:understanding_transport_DSF}. This differs from standard hydrodynamics where the heat conduction dampens the acoustic mode~\cite{boon1991molecular}, something also noted by Ref.~\cite{vieillefosse1975statistical}.

By comparing the generalised Bohm-Gross dispersion~\cite{arista1984dielectric} and equation \eqref{eq:dipersion}, one identifies the ideal behaviour as
\begin{equation}
    c_{s,e}^2 \approx \langle \vec{v}_{e}^2 \rangle = \begin{cases}
        3 \boltz T_{0,e} / m_e \quad &k_B T \gg E_{F}\\
        \frac{3}{5} v_F^2 \quad &k_B T \ll E_{F}
    \end{cases},
    \label{eq:sound_speed}
\end{equation} 
where $\langle \vec{v}_{e}^2 \rangle$ is the averaged square electron velocity, $v_F$ is the Fermi-speed, $E_F$ the Fermi-energy and $\boltz$ is the Boltzmann constant. In the absence of inter-species collisions, the damping rates are proportional to $\vec{k}^2$ and likewise the width of the resonance features in $S_{ee}^{0}(\vec{k}, \omega)$, which is the behaviour predicted by Schmidt \textit{et al.}~\cite{schmidt2012quantum}. Therefore, inter-species collisions are needed for the appropriate long-wavelength behaviour for two-component systems which typically have a finite width plasmon feature~\cite{glenzer2007observations,neumayer2010plasmons,fortmann2010influence}.

The effect of momentum transfer between the electrons and ions is the most clear when $\vec{k} \rightarrow 0$ and $g = 0$. In this limit, the dynamic structure factor \eqref{eq:hydro_DSF} yields, 
\begin{equation}
    \frac{\pi S^{0}_{ee}(\vec{k}, \omega)}{S^{0}_{ee}(\vec{k})} = \frac{\real\{\nu_{ei}\}\, \omega_p^2}{\left(\omega^2 - \omega_p^2 - \imaginary\{\nu_{ei}\}\omega \right)^2\! +\! \left(\real\{\nu_{ei}\}\omega\right)^2}
    \label{eq:DSF_hydro_BMA}
\end{equation}
where $\real\{ \nu_{ei} \}$ and $\imaginary\{ \nu_{ei} \}$ are the real and imaginary parts of the collision frequency respectively. This limit agrees with the Drude-like limit of Mermin's relaxation time approximation~\cite{fortmann2010influence}.

Lastly, one-component systems have been extensively studied through \textit{dynamic local field corrections} $G_{ee}(\vec{k}, \omega)$~\cite{farid1993extremal,moroni1995static,dornheim2019static}, which are related to the viscosity of the electron fluid~\cite{nifosi1998dynamic,conti1999elasticity}. In the long-wavelength limit $G_{ee}$ approaches zero, specifically $G_{ee}(\vec{k}, \omega) \propto \vec{k}^2$~\cite{fortmann2010influence}, and therefore, for the one-component case, the hydrodynamic limit of $S_{ee}^{0}$ is the Random Phase Approximation (RPA)~\cite{fortmann2010influence}, which is reflected in the hydrodynamic model when $\mathfrak{v}_{e} \rightarrow 0$ and $\mathfrak{h}_{e} \rightarrow 0$.

\subsection{Electron electrical conductivity}\label{sec:conductivity}
A commonly discussed property for the electrons not explicitly shown in the hydrodynamic formulation is the electrical conductivity, which relates to the system response under an external potential $\delta \phi_{\text{ext}}$. Introducing an external potential through $\phi \rightarrow \phi + \delta \phi_{\text{ext}}$, which is assumed to be small, a driving term 
\begin{equation}
    \Tilde{\vec{d}} = \Big[ 0,\; \frac{q_e \rho_{e,0}}{m_e}\vec{k}^2 \delta \Tilde{\phi}^{\text{ext}}_{\vec{k}},\; 0 \Big]^{\intercal}
\end{equation}
is added to the right-hand side of equation \eqref{eq:FL_lin}. On thermal averaging, initial perturbations vanish and the dielectric function $\varepsilon(\vec{k}, \omega)$ can be obtained from the density response~\cite{sturm1993dynamic} of the hydrodynamic model, and yields:
\begin{equation}
    \frac{1}{\varepsilon(\vec{k}, \omega)} = 1 + \frac{q_e^2 \rho_{0,e}}{\varepsilon_0 m_e^2} \left(s \mat{I} + \mat{D}_e(\vec{k}) \right)^{-1}_{12}\Big|_{s = -i\omega}.
    \label{eq:dielectric}
\end{equation}
The dielectric response relates to the off-diagonal component of the hydrodynamic matrix, as the density respond (row 1) to a source in the momentum equation (column 2). The resulting macroscopic conductivity model~\cite{jackson1975classical},
\begin{equation}
    \begin{aligned}
        \sigma(\omega) &= \varepsilon_0 \lim_{\vec{k} \rightarrow 0} \omega \Im\left\{ \varepsilon(\vec{k}, \omega) \right\}\\
        &= \frac{\varepsilon_0 \real\{ \nu_{ei} \} \omega_p^2}{\real\{ \nu_{ei} \}^2 + \left( \omega - \imaginary\{ \nu_{ei} \} \right)^2}
    \end{aligned}
    \label{eq:conductivity_model}
\end{equation}
has a Drude like form with DC value $\sigma_{\text {DC}} = \varepsilon_0 \omega_p^2 / \real\{ \nu_{ei}(\omega = 0) \}$. Electron-electron collisions do not appear explicitly as they adhere to the conservation relations in equation \eqref{eq:fluid}, however, they affect the conductivity by altering the distribution function~\cite{cohen1950electrical,spitzer1953transport}, which in a hydrodynamic formulation modifies the effective momentum transfer between electrons and ions $\nu_{ei}$~\cite{cauble1987transport}. 

Note that equation \eqref{eq:dielectric} constitutes a separate derivation of the dynamic structure factor using the (classical) fluctuation-dissipation theorem~\cite{kubo1966fluctuation,ichimaru1985theory,sturm1993dynamic}, which is consistent with the hydrodynamic form \eqref{eq:hydro_DSF} given the static structure factor
\begin{equation}
    S^{0}_{ee}(\vec{k}) = \frac{\boltz T}{m_e \omega_p^2}  \frac{\vec{k}^2}{1 + \lambda_e^{2} \vec{k}^2}
\end{equation}
and $\lambda_e^{-2} = \gamma_e \omega_p^2 / c_{s,e}^2$. This is in agreement with the static structure in the RPA approximation~\cite{ichimaru1985theory} for a general screening length $\lambda_e$ and recovers Debye and Thomas-Fermi screening~\cite{reinholz2005dielectric} in their corresponding limits, see appendix \ref{app:ref_values:ideal}. 

\section{Test system and model requirements}\label{sec:system}
The transport coefficients in the hydrodynamic model remain unspecified and will be extracted via comparison with molecular dynamics, where individual particle trajectories are resolved and no concrete concept of transport coefficients is needed. As we are interested in the electron dynamics, the WPMD method~\cite{klakow1994hydrogen,feldmeier2000molecular,grabowski2014review} is considered, specifically we employ the model presented in Ref.~\cite{svensson2023development}. For an initial test system we take a fully ionised hydrogen plasma with density $r_s = \left(4\pi m_e^{-1} \rho_e /3\right)^{-\frac{1}{3}} / \bohr = 2.0$ ($\bohr$ is the Bohr radius) and temperature $\boltz T \approx 21.5\,\text{eV}$. Further details of the MD model and the properties of the test system are provided in paper I.

\begin{figure}
    \centering
    \includegraphics[width=\linewidth]{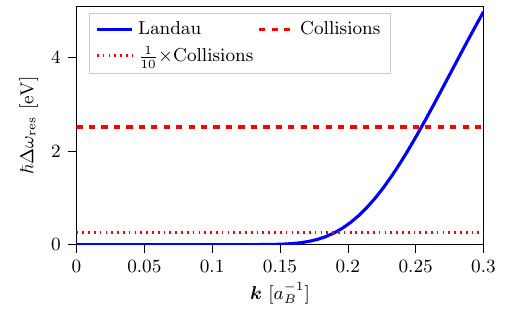}
    \caption{Width of the plasmon feature (FWHM) based on classical Landau damping~\cite{haas2011quantum} and the $\vec{k} \rightarrow 0$ limit~\cite{fortmann2010influence} of the Born-Mermin approximation~\cite{ropke1999lindhard} based on electron-ion collisions within the first Born approximation~\cite{thiele2008plasmon}. Ion structure taken form paper I. A tenth of the collisional width is shown for comparison with the Landau damping.} 
    \label{fig:requierd_k}
\end{figure}

The hydrodynamic description is valid as $\vec{k} \rightarrow 0$ and before attempting to apply it to the MD data an upper limit for $\vec{k}$ must be determined. A purely kinetic effect missing in the fluid model is Landau damping, which will act to broaden the resonance. Figure \ref{fig:requierd_k} compares the classical Landau damping~\cite{haas2011quantum} with the expected broadening due to electron-ion collisions in the first Born approximation~\cite{thiele2008plasmon} within the Mermin-relaxation time approximation~\cite{ropke1999lindhard}. Requiring the collisional broadening to be an order of magnitude larger than Landau damping, restricts the domain of validity to $\vec{k} \lesssim 0.19\,\bohr^{-1}$.

The required system size to model in the MD is set by the number of needed spectra within the regime of validity of the hydrodynamic model. Requiring access to at least three distinct spectra with $\vec{k} \leq 0.19\,\bohr^{-1}$ for the inference procedure and assuming MD simulations performed in a cubic box with periodic boundary conditions, at least $N_{\text{ion}} \approx 5600$ ions are needed. In combination with the necessity to resolve electron dynamics, a method like WPMD is therefore preferable compared to other \textit{ab initio} approaches. Here a system of $N_{\text{ion}} = 5628$ has been simulated.

A commonly discussed generalisation to the hydrodynamic formulation is quantum hydrodynamics (QHD), where an additional Bohm-like contribution is added to the momentum~\cite{haas2011quantum,khan2014quantum,moldabekov2018theoretical,bonitz2019quantum} and potentially higher order equations~\cite{haas2011quantum,schmidt2012quantum}. Classical hydrodynamics can still capture some quantum effects by modifying the equation of state and transport coefficients, but it can't capture the non-local Bohm contribution. We have opted for a more classical formulation for two main reasons. Quantum hydrodynamics adds (in quadrature) a quantum recoil $\omega_R = \hbar\vec{k}^2/(2m_e)$ to the plasmon dispersion~\cite{haas2011quantum,khan2014quantum}. Comparing the shift in the dispersion due to quantum recoil and the pressure contribution for the test case, 
\begin{equation}
    \frac{\omega_R^2}{c_{s,e}^2\vec{k}^2} \approx 0.1 \times \left(\bohr \vec{k}\right)^2,
    \label{eq:recoil_shift}
\end{equation}
it is clear that for our $\vec{k}$-vectors this contribution is negligible. In the above expression, the high-temperature approximation for $c_{s,e}^2$ has been assumed. Secondly, care must be taken for the Bohm contribution to have the desired effect in many-body systems~\cite{moldabekov2022towards}, partially amended by the introduction of an additional pre-factor for the Bohm potential~\cite{michta2015quantum} which might be time- and length-scale dependent~\cite{moldabekov2018theoretical}. Furthermore, the Bohm pressure tensor is necessarily anisotropic in the presence of perturbations, preventing a straightforward isotropic second-order closure~\cite{haas2011quantum}, i.e. closing on the level of temperature as in section \ref{sec:hydro}. Not properly accounted for this results in an erroneous recoil term~\cite{zhou19952}. Moreover, closing the equations on the level of the momentum equation would miss the entropy modes at $\omega = 0$~\cite{boon1991molecular,pines2018theory}.

\section{Fitting model, degeneracy and error estimation}\label{sec:MCMC}
It is common practice to fit hydrodynamic expressions to the simulated ion dynamic structure factor to extract transport properties, e.g. Refs.~\cite{mithen2011extent,kahlert2020thermodynamic,schorner2022extending,svensson2024modelling}, closely related to how it would be done using experimental data~\cite{mabey2016study,white2024speed}. Multiple $\vec{k}$-vectors (angles) are commonly needed to extract the transport properties. What we propose here is similar, however, here we are interested in the electron dynamics. 

For the ionic properties, a two-stage fitting procedure is commonly employed, where a hydrodynamic model is fitted for each $\vec{k}$ and the parameters in the fit are plotted as functions of $\vec{k}$ to extract transport properties~\cite{mithen2011extent,kahlert2020thermodynamic,schorner2022extending,svensson2024modelling}. A similar procedure using equation \eqref{eq:hydro_DSF} could also be implemented in our case, however, we are limited to fewer $\vec{k}$-vectors than in the ion dynamics investigations. It is therefore desirable to have a combined fit where uncertainties are systematically propagated. 

For this, we turn to Bayesian inference~\cite{von2011bayesian}, which is now often applied to scattering data for warm dense matter systems~\cite{clerouin2016bayesian,kasim2019inverse,poole2022case,poole2024multimessenger}. The fitting model is based on \textit{Bayes' theorem}, 
\begin{equation}
    P\left( \vec{x}| \vec{y} \right) = \frac{P\left( \vec{y} | \vec{x} \right) P\left( \vec{x} \right)}{P\left( \vec{y} \right)},
\end{equation}
for a set of transport coefficients and error parameters $\vec{x} = [c_{s,e}, \chi_{e}, \nu_{ei}, \nu_{L,e}, b_{\vec{k}_1}, s_{\vec{k}_1}, \dots, b_{\vec{k}_n}, s_{\vec{k}_n}]^{\intercal}$ and a set of observed (normalised) structure factors 
\begin{equation}
    \begin{aligned}
        \vec{y} &= [S(\vec{k}_1, \vec{\omega}_{\vec{k}_1}) / S(\vec{k}_1), \dots, S(\vec{k}_n, \vec{\omega}_{\vec{k}_n}) / S(\vec{k})]^{\intercal}\\
        &\equiv [y_1(\vec{\omega}_{\vec{k}_1}), \dots, y_n(\vec{\omega}_{\vec{k}_n})]^{\intercal},
    \end{aligned}
\end{equation}
where $S(\vec{k}, \omega)$ was obtained via a FFT of correlation data from MD. The effect of the finite time window in the MD was investigated in terms of window functions~\cite{prabhu2014window} and had a negligible effect. $P\left( \vec{x}| \vec{y} \right)$ is the resulting distribution of transport properties (and error parameters) given the data. The likelihood function $P\left( \vec{y} | \vec{x} \right)$ is a model choice. We consider the computations for different $\vec{k}$ to be independent measurements and each frequency point is modelled with a gaussian error around the hydrodynamic model $y_i^{\text{HD}}$ (equation \eqref{eq:hydro_DSF}),
\begin{equation}
    \begin{aligned}
        \ln P\left(\vec{y}| \vec{x}\right) = -\hspace{-0.1cm}\sum_{i=1}^n&\sum_{\omega \in \vec{\omega}_{\vec{k}_i}} \Bigg\{ \ln\left[2\pi \sigma_{i}^{2}(\omega) \right]\\
        &+ \frac{1}{2}\left[ \frac{ y_i(\omega) - y_i^{\text{HD}}(\omega; \vec{x})}{\sigma_{i}(\omega)}\right]^{2} \Bigg\}.
    \end{aligned}
\end{equation}
The  error is modelled as a floored relative error,
\begin{equation}
    \sigma_{i}(\omega) = \min\left[s_{i}, b_{i}y_i^{\text{HD}}(\omega) \right],
\end{equation}
where $s_{i}$ and $b_{i}$ are allowed to be different for each $\vec{k}$-vector and are inferred from the data.

The prior distribution $P\left(\vec{x}\right)$ for transport properties is assumed to be constant for positive values. \textit{Jefferys's prior}~\cite{von2011bayesian} was adopted for the scale parameters $s_i$ and $b_i$, based on the Fisher information matrix of a multivariate gaussian~\cite{malago2015information}. The likelihood for the observation $P\left( \vec{y} \right)$ will not be needed for parameter estimation~\cite{von2011bayesian}.

The resulting distribution, $P(\vec{x}|\vec{y})$, will be sampled numerically in the following section, using a Markov Chain Monte Carlo (MCMC) algorithm~\cite{metropolis1953equation,hastings1970monte}. For this, the \texttt{emcee} implementation~\cite{foreman2013emcee} of the affine-invariant ensemble sampler by Goodman and Weare~\cite{goodman2010ensemble} was employed.

The energy relaxation rate $\mathfrak{g}_e$ is a slow process compared to the damping of plasma oscillations, typically $\mathfrak{g}_e / \nu_{ei} \approx m_e / m_i$~\cite{rightley2021kinetic}, and therefore the free electron spectrum is not particularly sensitive to it and $\mathfrak{g}_e$ is omitted from the fitting. To further constrain the fitting procedure the adiabatic index is assumed to be $\gamma_e = 3$ -- common practice for fluid models of fast plasma oscillations~\cite{palastro2009kinetic} -- and a parameterisation for the dynamic collision frequency is introduced:
\begin{equation}
    \nu_{ei}(\omega) = \real\{ \nu_{ei}(\omega_p) \} \times \frac{\nu_{ei}^{\text{BMA}}(\omega)}{\real\{ \nu_{ei}^{\text{BMA}}(\omega_p) \}},
    \label{eq:collision_freq_param}
\end{equation}
where $\real\{ \nu_{ei}(\omega_p) \}$ will be sampled and $\nu_{ei}^{\text{BMA}}(\omega)$ is the collision frequency in the Born-Mermin approximation evaluated as described in Ref.~\cite{thiele2008plasmon} using the ion structure from paper I.

An alternative route for computing transport properties is from the Green-Kubo relations~\cite{boon1991molecular,hansen1993theory,grabowski2020review,stanek2024review} where the transport properties are related to integrals of auto-correlation functions of corresponding currents. However, they are derived from solutions of equations similar to \eqref{eq:FL_lin}. For ionic systems good agreement has been found between the Green-Kubo and hydrodynamic approach~\cite{mithen2012molecular} and the methods are largely interchangeable. A benefit of the hydrodynamic method is the validation of the model via inspection of the fits and its direct applicability to experimental data.

\section{Inference: Transport properties}\label{sec:inference}
\begin{figure*}
    \centering
    \includegraphics[width=\linewidth]{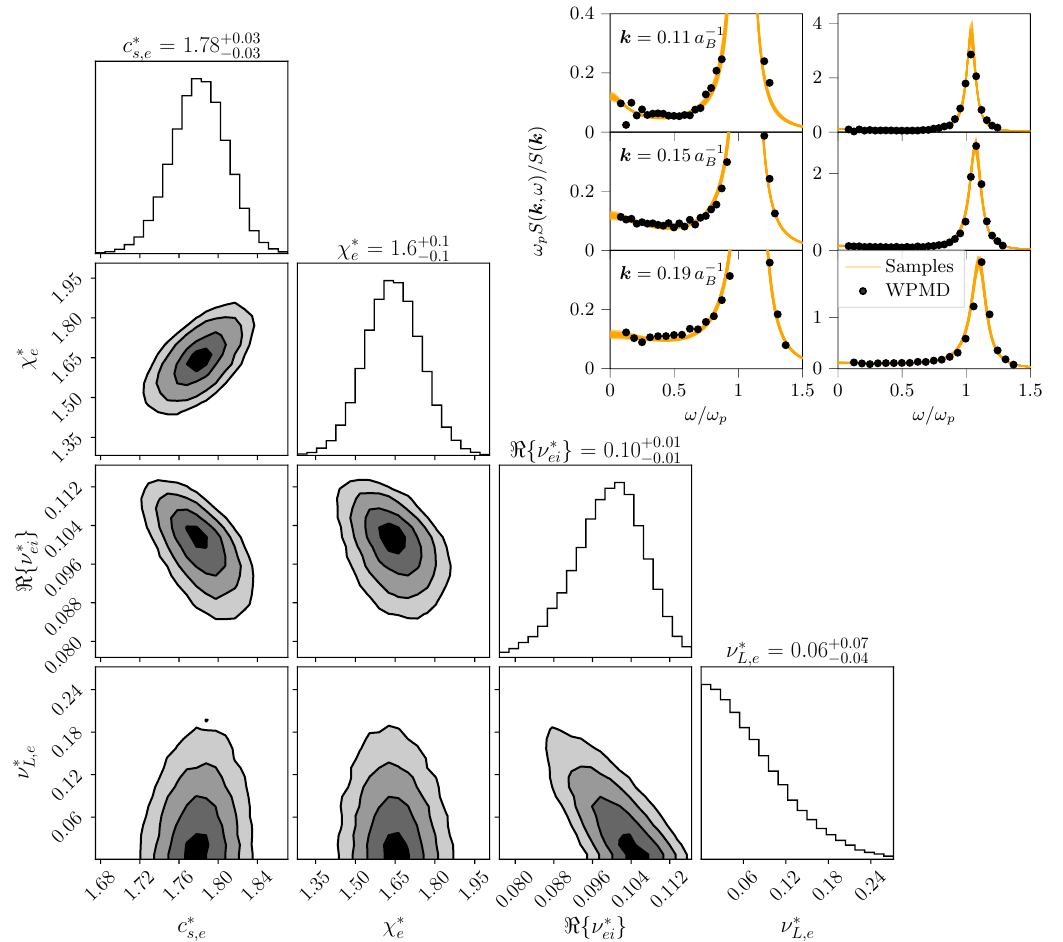}
    \caption{(Lower left) Statistics on the transport parameters as extracted from the MCMC sampling of the fitting model using the three smallest $\vec{k}$-modes in the WPMD. The collision frequency is to be understood as evaluated at $\omega = \omega_p$. Dimensionless units are used -- as described by equation \eqref{eq:dimless_transport_coeff} -- and the quoted values are the median. The errors are computed from the 16th and 84th percentiles. Figure generated using Ref.~\cite{corner}. (Upper right) Sample fits from the MCMC samples along with the MD data (WPMD). Each row corresponds to different $\vec{k}$-vectors as given in the figure. The right column shows the fitting of the plasma feature while the left columns focus on the entropy feature. The fitting model can fit both the plasmon and entropy features simultaneously.}
    \label{fig:transport_corner}
\end{figure*}
The Bayesian model has been used to simultaneously sample $S_{ee}^{0}(\vec{k}, \omega)$ for the three smallest $\vec{k}$-vectors in the MD simulation. The resulting posterior distributions for the transport parameters as well as sample fits are shown in figure \ref{fig:transport_corner}. Transport parameters are given in dimensionless units,
\begin{equation}
    \begin{aligned}[c]
        c_{s,e}^{*} &= \sqrt{\frac{m_e}{\boltz T_{e,0}}} c_{e,s}\\
        \chi_{e}^{*} &= \sqrt{\frac{m_e}{a_e^2 \boltz T_{e,0}}} \chi_{e}
    \end{aligned}
    \hspace{0.5cm}
    \begin{aligned}[c]
        \nu_{ei}^{*} &= \sqrt{\frac{m_e a_e^2}{\boltz T_{e,0}}} \nu_{ei}\\
        \nu_{L,e}^{*} &= \sqrt{\frac{m_e}{a_e^2 \boltz T_{e,0}}} \nu_{L,e},
    \end{aligned}
    \label{eq:dimless_transport_coeff}
\end{equation}
where $a_e = r_s \bohr$. The sample fits in figure \ref{fig:transport_corner} demonstrate the ability of the hydrodynamic model to fit both the plasmon feature and the entropy feature at $\omega = 0$ for multiple $\vec{k}$ simultaneously. This strengthens the confidence in the model's applicability for long wavelengths and frequencies up to the plasma frequency.

The inferred parameter distributions show correlations, particularly between the collision frequency $\real\{\nu_{ei}(\omega_p)\}$ and the viscosity $\nu_{L,e}$, because they both relate to the width of the plasmon feature. However, the inferred distribution for the viscosity does not show a clear peak away from the boundary $\nu_{L,e} = 0$, suggesting this $\vec{k}$-dependence of the plasmon feature is challenging to measure at small $\vec{k}$. Other correlation functions, rather than the density-density considered here, might be preferable to further probe the viscosity. Furthermore, correlations between $c_{s,e}$ and $\nu_{ei}$ represent the uncertainty between the width of the plasmon feature and its exact position in the discrete data, and the correlation between $c_{s,e}$ and $\chi_e$ demonstrates the sensitivity of the comparatively small entropy feature to the ''wings'' of the plasmon feature.

\begin{table*}[]
    \centering
    \caption{Extracted and reference transport properties for the electron fluid. Values are in dimensionless form as given by equations \eqref{eq:dimless_transport_coeff} and \eqref{eq:dimless_sigma}. The inferred values for the transport coefficients are based on the fitting procedure discussed in section \ref{sec:MCMC} for the two (not shown) and three (see figure \ref{fig:transport_corner}) smallest $\vec{k}$-vectors in the molecular dynamics. Values are the median and the errors are computed from the 16th and 84th percentiles. Reference calculations are discussed in appendix \ref{app:ref_values}.}
    \label{tab:reference_transport}
    \def\fixupspeed{-1.6em}
    \def\fixupthermal{-1.0em}
    \def\fixupviscosity{-1.0em}
    \def\fixupconductivity{-0.5em}
    \begin{ruledtabular}
    \begin{tabular}{c|cc|ccc}

        & \multicolumn{2}{c|}{WPMD \& MCMC} & \multicolumn{3}{c}{Reference calculations}\\
        
         Property & \#$\vec{k} = 2$ & \#$\vec{k} = 3$ & Value & Comment & Section \\\hline

         \multirow{2}{*}[\fixupspeed]{$c_{s,e}^{*}$} & \multirow{2}{*}[\fixupspeed]{$1.74^{+0.04}_{-0.04}$} & \multirow{2}{*}[\fixupspeed]{$1.78^{+0.03}_{-0.03}$} & $\sqrt{3}$ & \makecell{Bohm-Gross dispersion in high temperature\\ approximation, equation \eqref{eq:sound_speed}.} & \ref{app:ref_values:ideal}\\
         
          & & & $1.69$ & \makecell{Memory function and parameterisation of Monte Carlo\\ simulations for the one component plasma.} & \ref{app:ref_values:OCP}\\

          & & & $1.78$ & \makecell{Modified Bohm-Gross dispersion based on the $\vec{k} \rightarrow 0$\\ limit of RPA for moderately degeneracy.} & \ref{app:ref_values:born}\\\cline{4-6}

         \multirow{3}{*}[\fixupthermal]{$\chi_e^{*}$} & \multirow{3}{*}[\fixupthermal]{$1.6^{+0.1}_{-0.1}$} & \multirow{3}{*}[\fixupthermal]{$1.6^{+0.1}_{-0.1}$} & $2.3$ & \makecell{Classical kinetic SMT model with effective Boltzmann\\ formulation for collisions and ideal heat capacity.} & \ref{app:ref_values:ideal} \& \ref{app:ref_values:smt}\\
         
          & & & $2.5$ & \makecell{Parameterisation of (classical) MD and Monte Carlo\\ simulations for one component plasma.} & \ref{app:ref_values:OCP}\\
          
          & & & $2.8$ & \makecell{Parameterisation of Coulomb log for thermal\\ conductivity by DFT-MD and ideal heat capacity.} & \ref{app:ref_values:ideal} \& \ref{app:ref_values:DFTMD} \\\cline{4-6}

         $\real\{\nu_{ei}^{*}(\omega_p)\}$ & $0.10^{+0.01}_{-0.01}$ & $0.10^{+0.01}_{-0.01}$ & $0.16$\footnote{\label{fot:collision_and_conductivity}The collision frequency and conductivity is different representation of the same property within the model, and therefore the same models are used for comparison evaluated at the appropriate $\omega$.} & \makecell{Dynamic collision frequency at $\omega = \omega_p$ in first Born\\ approximation.} & \ref{app:ref_values:born}\\\cline{4-6}

        \multirow{3}{*}[\fixupviscosity]{$\nu_{L,e}^{*}$} & \multirow{3}{*}[\fixupviscosity]{$0.07^{+0.09}_{-0.05}$} & \multirow{3}{*}[\fixupviscosity]{$0.06^{+0.07}_{-0.04}$} & $7.5$ & \makecell{Classical kinetic SMT model with effective Boltzmann\\ formulation for collisions.} & \ref{app:ref_values:smt}\\

        & & & $4.6$ & \makecell{Parameterisation of (classical) MD simulations for\\ one component plasma.} & \ref{app:ref_values:OCP}\\
        
        & & & $8.7 \times 10^{-3}$ & \makecell{$T = 0$ parametrisation for a homogeneous quantum\\ Fermi liquid.} & \ref{app:ref_values:conti}\\\hline

        \multirow{2}{*}[\fixupconductivity]{$\sigma_{\text{DC}}^{*}$} & \multirow{2}{*}[\fixupconductivity]{$16.1^{+1.6}_{-1.2}$} & \multirow{2}{*}[\fixupconductivity]{$16.7^{+1.4}_{-1.1}$} & $12.0$ & \makecell{Classical kinetic SMT model with effective Boltzmann\\ formulation for collisions.} & \ref{app:ref_values:smt}\\

        & & & $10.4$$^{\text{\ref{fot:collision_and_conductivity}}}$ & \makecell{Based on dynamic collision frequency at $\omega = 0$ in first Born\\ approximation and the conductivity model of eq.~\eqref{eq:conductivity_model}.} & \ref{app:ref_values:born}

    \end{tabular}
    \end{ruledtabular}
\end{table*}

The inferred electron transport coefficients with error estimates are summarised in table \ref{tab:reference_transport} along with an analogous sampling where only the two spectra are fitted (not shown). Using the smaller number of spectra yields transport coefficients within the error bands of the original fits, although with larger uncertainty. This is indicative of the MD simulations having reached a length scale well described by the hydrodynamic formulation. This is further supported if additional spectra with $\vec{k} \gtrsim 0.24\,\bohr^{-1}$ are included in the fit (not shown), which show an erroneous attribution of plasmon broadening due to Landau damping to viscous damping. The length scale at which this occur agrees with the estimates in section \ref{sec:system}.

Along the inferred values from the WPMD simulations in table \ref{tab:reference_transport}, a selection of reference calculations are presented for the transport properties. Details of the reference calculations are provided in appendix \ref{app:ref_values}. We notice that $c_{s,e}$ is approximately $3\%$ higher than the ideal value given by equation \eqref{eq:sound_speed}. The discrepancy might be attributed to quantum corrections as no strong deviations are expected from coulomb coupling (one-component plasma comparison), and calculations based on the RPA increase $c_{s,e}$ in agreement with the present results. However, based on the estimated errors, we can't conclusively distinguish the models.

The thermal diffusivity is the ratio of the thermal conductivity -- which has been tabulated in a variety of different models -- and the heat capacity. In comparison with classical kinetic models and MD data for a one component plasma, the WPMD fit suggest a $35\%$ lower thermal diffusivity. A quantum two component calculation using DFT-MD instead suggests a $75\%$ higher value than the WPMD fits, however, this reference model uses an ideal heat capacity whereas the inclusion of interactions would reduce the thermal diffusivity. 

The collision frequency for momentum loss at the plasma frequency $\real\{ \nu_{ei}^{*}(\omega_p) \}$ is comparable to the values given by the first Born approximation albeit a bit lower. This is possibly related to the distributed form of the electron wave-packet which will reduce the strength of collisions.

Lastly, whereas the previously discussed inferred properties have largely been in agreement with classical models, the inferred electron viscosity is markedly lower than the classical counterparts. The quoted values for the longitudinal viscosity $\nu_{L,e}$ in all reference calculations have neglected the contribution from bulk viscosity $\zeta_e$, which both in a quantum zero temperature~\cite{nifosi1998dynamic,conti1999elasticity} and classical~\cite{vieillefosse1975statistical} one component plasma have been shown to be negligible compared to the shear viscosity $\eta_{e}$. Due to the distribution of sampled $\nu_{L,e}$, we can confidently estimate an order of magnitude reduction compared to the classical model. However, the predicted value is still above the $T = 0$ limit.

\begin{figure}
    \centering
    \includegraphics[width=\linewidth]{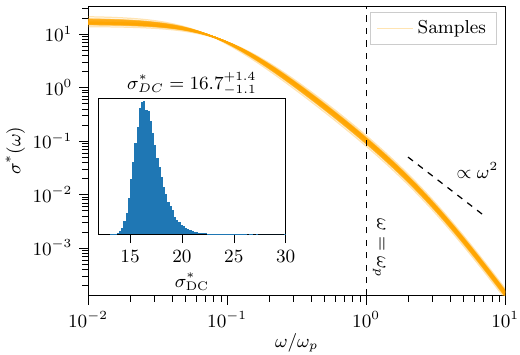}
    \caption{The conductivity model evaluated based on samples of the transport coefficients from the MCMC. The model is seen with a Drude like character where the high-frequency behaviour decays as $\omega^{-2}$, however, for $\omega > \omega_p$ the frequency dependence on $\nu_{ei}$ introduces an additional suppression. In the insert the distribution of DC conductivities, $\sigma^{*}_{\text{DC}}$, is shown. Quoted value is the median and the error is estimated from the 16th and 84th percentiles.}
    \label{fig:conductivity}
\end{figure}

Given the inferred values for $\nu_{ei}$, the conductivity model in section \ref{sec:conductivity} is evaluated in figure \ref{fig:conductivity}, using dimensionless units 
\begin{equation}
    \sigma^{*}(\omega) = \sqrt{\frac{m_e a_e^2}{\boltz T_{e,0} \varepsilon_0^2}} \sigma(\omega).
    \label{eq:dimless_sigma}
\end{equation}
The conductivity model has a Drude-like form, where the additional suppression for large $\omega$ comes from the frequency dependence of the collision frequency. The inferred collision frequency is sensitive to the spectrum close to $\omega \approx \omega_p$, and the high-frequency behaviour is extrapolated based on the model of dynamic collision frequencies. Similar extrapolation is performed for the obtained DC conductivity -- shown in the inset of figure \ref{fig:conductivity} -- in this case the static and $\omega = \omega_p$ collision frequencies differ by $15\%$. Recently, Hentschel \textit{et~al.}~\cite{hentschel2024statistical} discussed the sensitivity of extrapolating a measurement of the dynamic collision frequency at $\omega \approx \omega_p$ to the DC conductivity, where the inference becomes uncertain as more free parameters are included in the model for the dynamic collision frequency. The comparison of DC conductivities, see table \ref{tab:reference_transport}, reaches the same conclusion as for the collision frequency in general.

\section{Conclusion}\label{sec:conclusion}
In an effort to describe the long-wavelength behaviour of atomistic two component simulations of warm dense matter systems, a phenomenological two fluid model is considered. In particular, the hydrodynamic model includes self-consistent electrostatic fields and collisional effects, and models ion screening on the ion time scales and plasma oscillations on the electron time scales. 

The dynamic structure factor for free-electron oscillations is derived both by considering the dynamics of density fluctuations and the response of the systems due to external potentials, via linearisation of the hydrodynamic model. In taking the appropriate limits, it is shown explicitly that the model incorporates one entropy and two plasmon modes, and agrees with the Born-Mermin approximation in the long-wavelength limit. It is therefore an appropriate description of the electron dynamics with ion collisions on long length scales.

The dynamic structure factor is parameterised using transport parameters, and therefore, by mapping the structure factor to the results of atomistic simulations, transport properties can be extracted. The benefit here is the direct and clear interpretation of the transport coefficients in the original hydrodynamic formulation. We perform the mapping process via Markov Chain Monte Carlo (MCMC) sampling, over multiple spectra simultaneously, which allows for consistent error propagation.

The methodology is general and can be applied to any simulation of two-component plasmas in which the electron dynamics is probed. In this work, we have applied it to the results from wave packet molecular dynamics (WPMD) of warm dense hydrogen. Fractionally small error margins in the transport properties are obtained, and the inferred values are in agreement with a number of reference calculations. Discrepancies are within expectations as no reference calculation is a direct match to the WPMD system, either due to being limited to small coupling or having neglected the presence of ions. The exception is the electron viscosity, which, when compared to classical models, is suggested to be at least an order of magnitude smaller. This discrepancy is most likely due to quantum effects in the electron-electron interaction not present in the classical references. 

\begin{acknowledgments}
    We acknowledge stimulating discussions with H.~Poole and C.~Heaton.
    The authors are gratefully for the use of computing resources provided at STFC Scientific Computing Department’s SCARF cluster, where the wave packet computations has been performed. PS acknowledges funding from the Oxford Physics Endowment for Graduates (OXPEG). PS, DP, SMV and GG acknowledge support from AWE-NST via the Oxford Centre for High Energy Density Science (OxCHEDS). The work of GG and SMV has received partial support from EPSRC and First Light Fusion under AMPLIFI Prosperity Partnership grant no.\ EP/X25373/1. SMV acknowledges support from EPSRC grant EP/W010087/1.
\end{acknowledgments}

\appendix

\section{Reference transport properties}\label{app:ref_values}
In section \ref{sec:inference} the transport properties of the WPMD model are inferred via Bayesian parameter estimation. Along with these results, a set of reference calculations are shown for comparison. In this appendix, more details on the reference computations are provided.

\subsection{Ideal treatment}\label{app:ref_values:ideal}
The ideal sound speed is given by comparison to the generalised Bohm-Gross dispersion~\cite{arista1984dielectric} in equation \eqref{eq:sound_speed}. The ideal pressure at constant temperature, $p_e = \rho_{e} \left\langle \vec{v}^2 \right\rangle / 3$, and the thermodynamic relation~\cite{boon1991molecular},
\begin{equation}
    c_{s,e}^2 = \gamma_e \deri{p_e}{\rho_e}\Big|_{T_e}.
    \label{eq:sound_adiabatic_index}
\end{equation}
also provide a reference point. For equations \eqref{eq:sound_speed} and \eqref{eq:sound_adiabatic_index} to coincide, $\gamma_e = 3$ and $\gamma_e = 9/5$ is needed in the high- and low-temperature limits respectively. This differs from the thermodynamics for an ideal system, where $\gamma_e = 5/3$~\cite{landau2013statistical}, because the adiabatic index is commonly modelled as time scale dependent~\cite{palastro2009kinetic}.

To compute the thermal diffusivity, $\chi_e$ in some reference models the heat capacity $C_{V,e}$ is required. Within an ideal treatment and the high temperature approximation $\boltz T \gg E_F$, the heat capacity is given by $\rho_{0,e}C_{V,e} = 3 \boltz \rho_{e,0}/(2 m_e)$.

\subsection{Stanton and Murillo (2016)}\label{app:ref_values:smt}
The Stanton and Murillo transport (SMT) model is a classical kinetic transport model based on collisions in an effective Boltzmann description using Yukawa interactions~\cite{stanton2016ionic}. The model originally developed for ions has recently been applied to electron transport properties~\cite{stanton2021efficient}. Opposite charges can't be treated in the formulation and electrons are effectively treated as positrons, therefore this model does not adhere to the \textit{Barkas effect}~\cite{smith1953measurements,barkas1956mass}. Despite this and limited quantum effects, the model captures the general trend for a variety of transport properties~\cite{stanton2021efficient,stanek2024review}. 

The SMT model has been used for the computation of a variety of transport properties. For the viscosity and thermal conductivity a single-component formulation~\cite{stanton2016ionic} has been used, while the presence of ions is still included in the screening. We employ the two-component system formulation for the computation of the electrical conductivity~\cite{stanton2021efficient}.

\subsection{One component plasma}\label{app:ref_values:OCP}
The existence of a hydrodynamic description of a one component plasma is generally debated~\cite{baus1980statistical,mithen2011extent}, but the system is well studied numerically. Hansen~\cite{hansen1981plasmon} investigated the dispersion relation via a memory function formalism which has been tested against MD~\cite{mithen2012onset}. Here, the corrected dispersion has been evaluated using the excess energy from Monte Carlo simulations~\cite{caillol1999thermodynamic,caillol2010accurate} parameterised by Ref.~\cite{plummer2024ionisation}. 

Scheiner and Baalrud~\cite{scheiner2019testing} parameterised the thermal conductivity from MD simulations, which in combination with the heat capacity described by Hansen~\cite{hansen1973statistical} (and the excess energy parameterisation), has been used to evaluate the thermal diffusivity. Similarly, Daligault \textit{et al.}~\cite{daligault2014determination} provided a parameterisation of the shear viscosity valid for the gas and fluid regime, which is also taken for comparison.

\subsection{RPA and first Born approximation}\label{app:ref_values:born}
Within the long-wavelength limit of the RPA approximation Thiele \textit{et al.}~\cite{thiele2008plasmon} obtained a correction to the ideal dispersion for moderately degenerate plasmas. In our notation, the correction increases $c_{s,e}$ by a factor $\sqrt{1 + 0.088 \rho_e \Lambda_e^3 / m_e}$ from the ideal value, where $\Lambda_e$ is the thermal de Broglie wavelength. 

The model for the dynamic collision frequency in equation \eqref{eq:collision_freq_param} is compared to the first Born approximation~\cite{thiele2008plasmon} given the ionic structure from paper I. Evaluation at both  $\omega = \omega_p$ and $\omega = 0$, provide an estimate for both the sampled collision frequency and DC conductivity via equation \eqref{eq:conductivity_model}.

\subsection{DFT -- Molecular dynamics}\label{app:ref_values:DFTMD}
The electron thermal conductivity is parameterised via a Coulomb logarithm extracted from DFT-MD simulations of deuterium~\cite{hu2014first}. The simulations were performed at higher densities, where the lowest density considered, $r_s \approx 1.75$, is close to our test case, and the classical limit was regularised based on classical results. However, lacking an appropriate heat capacity, the ideal heat capacity is used which overestimates the thermal diffusivity.

\subsection{Conti and Vignale (1999)}\label{app:ref_values:conti}
Conti and Vignale~\cite{conti1999elasticity} provided a zero temperature parametrisation of the shear viscosity for a homogeneous quantum Fermi liquid, however, their result is almost three orders of magnitude lower than the classical kinetic results suggesting that finite temperature effects are dominant at the studied conditions.

\bibliography{ref}

\end{document}